\documentstyle[aps,preprint]{revtex}
\def\prl#1#2#3{{ Phys. Rev. Lett.} {\bf #1}, #2 (#3)}

\def\jsp#1#2#3{ J. Stat.  Phys. {\bf #1}, #2 (#3)}

\def\pre#1#2#3{Phys. Rev. E {\bf #1}, #2 (#3)}
\def\pra#1#2#3{Phys. Rev. A {\bf #1}, #2 (#3)}

\def\jpa#1#2#3{J. Phys. A {\bf #1}, #2 (#3)}

\def\physd#1#2#3{Physica D {\bf #1}, #2 (#3)}

\def\etc{ etc.~}
\def\X{\bf X}
\def\e{\bf e}

\def\etl{$et ~al.$~}

\def\lle{ local Lyapunov exponent }
\def\lles{ local Lyapunov exponents}
\def\le{ Lyapunov exponent~}

\def\L{\Lambda}

\def\local{\underline{local}~}
\def\convex{\underline{convex}~}

\def\beq{\begin{equation}}
\def\bc{\begin{center}}
\def\ec{\end{center}}
\def\eqn{\end{equation}}
\begin{document}
\title{Characteristic distributions of finite--time Lyapunov
exponents}
\author{Awadhesh Prasad and Ramakrishna Ramaswamy} 
\address{School of Physical Sciences\\ Jawaharlal Nehru University, New Delhi
110 067, INDIA}
\date{\today}
\maketitle
\begin{abstract}
We study the probability densities of finite--time or \local Lyapunov
exponents (LLEs) in low--dimensional chaotic systems. 
While the multifractal formalism describes how these densities
behave in the asymptotic or long--time limit, there are significant
finite--size corrections which are coordinate dependent.
Depending on the nature
of the dynamical state, the distribution of local Lyapunov exponents
has a characteristic shape. For intermittent dynamics, and at crises,
dynamical correlations lead to distributions with stretched exponential
tails, while for fully--developed chaos the probability density has a
cusp.  Exact results are presented for the logistic map, $x \to
4x(1-x)$.  At intermittency the density is markedly asymmetric, while
for `typical' chaos, it is known that the central limit theorem obtains
and a Gaussian density results. Local analysis provides
information on the variation of predictability on dynamical attractors.
These densities, which are used to characterize the {\sl nonuniform}
spatial organization on chaotic attractors are robust to noise and can
therefore be measured from experimental data.  
\end{abstract}

\section{INTRODUCTION}

The statistics of distributions of Lyapunov exponents (sometimes called
stretch--exponents) have been studied in a number of physical
situations ranging from turbulent flows \cite{tu1} to Hamiltonian
dynamics (in many--particle systems \cite{amitrano} and conservative
mappings \cite{sepulveda}), and are related to generalized dimensions
and entropies \cite{bs,gbp}.

Lyapunov exponents provide a quantitative characterization of dynamics:
for a dynamical system in $D$--dimensions, the $D$ Lyapunov exponents
of an orbit measure the rate at which volume elements in the phase
space expand or contract along the orbit, in the different directions.
A positive Lyapunov exponent (LE) signifies exponential divergence of
trajectories in the given direction, and is associated with chaotic
dynamics, while negative LEs are associated with stable motion, when
nearby trajectories converge.  Although the exponents are global or
asymptotic quantities, it is often instructive to examine the
distribution of values that the LE may take locally, namely over
finite-time segments along a given trajectory \cite{gbp,abar,ott}.  If
the underlying attractor is nonuniform, on a chaotic trajectory the
{\it local} LE can be negative within a finite time interval.
Similarly, on a nonchaotic trajectory the local LE can take positive
values over finite time intervals \cite{pf}.  These considerations are of
additional relevance for Lyapunov exponents computed from experimental
data \cite{wolf}, since these are, naturally, only finite--time
exponents. 

In this paper, we study the statistics of finite--time or local LEs in
low--dimensional dissipative dynamical systems. Local Lyapunov
exponents, which are defined in Eq.~(\ref{lle}) below, depend on
initial conditions, unlike the asymptotic or global Lyapunov exponent.
The distribution of values that the local exponents take depends, in a
characteristic manner, on the nature of the dynamical state; our
present focus is on characterizing the different distributions that
obtain for different dynamical attractors.

We find that the characteristic densities of Lyapunov exponents 
fall in distinct classes depending on the nature of the attractor.
Although the LLE distributions are stationary, they depend on the 
time interval over which the finite--time LE is computed. For very
short times, the distributions keep changing shape and are difficult
to classify, while in the asymptotic limit, as the time interval 
$\to \infty$, all distributions must eventually collapse to a
$\delta$--function centered on the global LE. 
The manner in which this happens is usually described by the multifractal
formalism \cite{gbp}, but here we address the  important corrections
to scaling that can obtain for finite--times.  

Some aspects of such distributions have been studied previously.
For  ``typical'' chaotic dynamics, when correlations die out 
exponentially rapidly, the central--limit theorem holds for 
a number of averaged quantities, including local Lyapunov 
exponents \cite{gbp,ott}. Thus, the density is a Gaussian 
function, whose width depends on the length of the time interval
over which the exponents are computed.  For intermittent systems on 
the other hand, it is well known that correlations die out very 
slowly; this can lead to a power--law scaling for several quantities 
such as the Lyapunov exponent or the diffusion constant 
\cite{fujisaka}. Benzi \etl also studied intermittency in 
more detail and observed a deviation from the normal distribution 
in quantities such as the fluctuations of the response function \cite{benzi}.

We extend the analysis of finite--time Lyapunov exponents to the 
particular case of fully developed chaos and intermittency, where 
dynamical correlations persist over long times. This leads to 
significant departures from  the simple central limit behaviour,
and the resulting distributions are quite distinct from the 
normal density, typically having exponential or stretched
exponential tails.

An example which can be solved exactly is the commonly studied 
logistic map at the Ulam point, namely $x \to 4 x (1 -x)$, for 
which we obtain an analytic form of the probability density for 
all times. The same distribution occurs for all parameter values 
where there is a boundary crisis, and thus appears to be quite 
general.

We also treat the case of intermittent dynamics in some detail. In all
instances of intermittency, the dynamics switches between two or more
distinct types of behaviour. The distribution of LLEs that arise in
such a situation can be shown to have components arising separately
from these individual behaviours.

Our main results are presented in Section II, where we 
discuss the different types of distributions for finite--time LEs.
Our studies have been mainly on simple dynamical systems such as
the logistic mapping, but the results we obtain appear to be 
more generally valid: these distributions can be seen a variety of 
systems (both mappings and flows).
This is followed by a summary and conclusions in Section III.

\section{Characteristic Densities of Finite--time exponents}

For generality, consider a $D$--dimensional  discrete nonlinear system 
\cite{foot1}
\begin{eqnarray}
\label{function} 
\X_{n+1} & = & F_{\{\alpha\}}(\X_n) 
\end{eqnarray}
\noindent
where $\X \in \hbox{R}^D$ and $\{\alpha \} $ is a set of parameters.
There are $D$ Lyapunov exponents which are defined by considering a 
set of orthonormal $D$-dimensional vectors  $\hat{\e}^m, m = 1,\ldots, D$, 
and examining their evolution under the effect of the tangent mapping which is
determined by the  Jacobian of $F$, namely $JF(\X)$.  
Defining 
\begin{equation} 
{\e}_j^m  =   (J F_{\{\alpha \}}({\bf X}_{j-1}),\hat{\e}_{j-1}^m),
\label{lyap}
\end{equation}\noindent
the Lyapunov exponents are  \beq
\Lambda^m = \lim_{N\to \infty} \frac{1}{N} \sum_{j=1}^N \ln \|{\e}_{j}^m \|,
~~~~~~~~ m = 1,2,\ldots D.
\label{lyap1}
\eqn
The vectors $\hat{\e}^m$ are re--orthonormalized along the trajectory, 
and the
subscript $j$ here refers to the time. Stretch exponents are the 
logarithms of the ratios by which the vectors expand 
(or contract) along the $D$ directions,
\beq
\lambda_1^m (j)  =  \ln \|{\e}_{j}^m \|,
\eqn
and this helps to define the $m$-th finite time Lyapunov exponent 
in a time interval of length $N$ as
\beq
\lambda_N^m  =  \frac{1}{N} \sum_{j=1}^N \ln \|{\e}_{j}^m \|
\eqn

In the remainder of this article only $m\equiv 1$, the 
largest Lyapunov exponent, 
is considered, so the superscript index will be omitted
henceforth in order to simplify notation.
Local Lyapunov exponents are calculated along a trajectory that
is divided into segments of length $N$.  By $\lambda_N(i)$ is meant the
$N$-step Lyapunov exponent calculated from the $i$th segment, and
clearly
\beq
\label{lle}
\lambda_N(k) = {1 \over N} \sum_{j = N(k-1)+1}^{kN} \lambda_1(j).
\eqn
The asymptotic Lyapunov exponent $\Lambda \equiv \lambda_{\infty}$ 
does not depend (with probability 1) on initial 
conditions but $\lambda_N$ does.  The 
{\it probability density} of local Lyapunov exponents,
which is a stationary quantity,  is defined as
\beq
P_{\alpha}( \lambda,N) d\lambda \equiv \mbox{ Probability that}
~\lambda_N
\mbox{~takes a value between} ~~\lambda \mbox{~and~} \lambda + d\lambda.
\eqn

If the stretch exponents of a system are considered to be random
variables since the dynamics is chaotic,
then the finite--time LEs should obey the central limit theorem, and
the distribution function can be written in the general form \cite{ott}
\beq
P(\lambda, N) \sim {1 \over [2\pi N \Phi^{\prime\prime}]^{1/2}}
\exp -N \Phi(\lambda),
\eqn
where $\Phi(\lambda)$ is a \convex function with 
minimum at $\lambda = \Lambda$.
Expanding $\Phi$ to second order gives the Gaussian density 
\beq
P_{\alpha}( \lambda,N)\sim \exp - (\lambda - \L)^2/2\sigma ,
\label{old}
\eqn 
with variance $\sigma^2 \propto 1/N $. 
Indeed this argument has been used very effectively to analyze
finite time LEs on so--called ``typical'' chaotic attractors 
\cite{gbp,ott,fujisaka} as shown in Fig.~1(a), where the Gaussian 
nature of the density is evident. However, for other dynamical 
attractors, $P(\lambda,N)$ can be
quite different, as shown in Figs.~1(b)--(d). Indeed, the 
departure of $\Phi(\lambda)$ from a polynomial function with 
quadratic maximum has been used to characterize the state and 
study the persistence of correlations \cite{sepulveda,mr,corre}.

Regardless of the behaviour of $P(\lambda,N)$ for small N, 
the local LEs eventually converge to the global exponent,
$\lim_{N \to \infty} P_{\alpha}(\lambda,N) \to \delta (\L -\lambda)$.  
However for sufficient large $N$ (but still far from the limit $N \to \infty$)
the characteristic distributions depend on the details of the dynamics,
so that the approach to the limit is distinctive. 
In the following subsections we discuss the particular cases of 
fully--developed chaos and intermittency in detail.

\subsection{Fully Developed Chaos}  The case of fully
developed chaos in a system such as the logistic mapping,
\beq x_{n+1} =  \alpha x_n (1 -x_n) \label{logistic} \eqn 
at $\alpha = 4$ can be analyzed in detail since the 
invariant density is known
exactly.  The Lyapunov exponent for this system is
$\Lambda = \ln 2$, and the invariant density, which can be 
obtained by solving  the appropriate Frobenius--Perron 
equation is \cite{intro,ce}
\beq
\rho(x) = 1/\pi\sqrt{ x(1-x)}, x \in~[0,1].
\label{density}
\eqn

The one--step Lyapunov exponent, 
\beq
\label{1step}
\lambda_1(i) \equiv \ln \vert \alpha (1-2x_i) \vert
\eqn
itself obeys the mapping
\beq
\lambda_1(i+1) = \ln \vert [\exp\{ 2 \lambda_1(i) \} - \alpha^2 + 2
\alpha ] /2 \vert 
\label{1st}
\eqn
$P(\lambda, 1)$ is merely the invariant density for this mapping, and 
by using Eq.~(\ref{density}), one obtains
\beq
\label{pl1}
P( \lambda,1)=\frac{2 \exp (\lambda - \ln 4)}{ \pi
\sqrt{1 - \exp [2(\lambda - \ln 4)]}}, ~~~~~-\infty 
\leq \lambda \leq \ln 4.
\eqn
$P(\lambda, N)$ can be calculated recursively since $\lambda_N$ is known in 
terms of $\lambda_1$ through Eq. (\ref{lle}).  Since $\lambda_1$
is known in terms of $x_i$ [Eq.~(\ref{1step})], it is possible
to reexpress Eq.~(\ref{lle}) as 
\beq
\exp (N\lambda_N) - G(x) = 0,\label{roots}
\eqn
where $G$ is a polynomial function of $x$ of order 2$^N$-1.
From this it directly follows that
\beq
\label{result}
P( \lambda,N) = N \exp(N \lambda) \sum_{\rm roots}\frac{\rho(x)} {\vert
G^{\prime}(x)\vert}
\eqn
the sum being over all real roots, $x(\lambda)$, of Eq.~(\ref{roots}) 
at given $\lambda \in [-\infty, \ln 4]$. Since the polynomial $G(x)$
is of odd order, there is always one real root.  For sufficiently small
$\lambda$ all roots are real; with increasing $\lambda$ they leave the
real axis in pairs, each such point giving rise to a singularity since
the derivative $dG(x)/dx$ vanishes there.

In principle, Eq.~(\ref{result}) provides an exact solution for
the invariant density of finite--time Lyapunov exponents for the logistic
mapping. Furthermore, the same technique can be applied to obtain 
invariant densities for other systems if $\rho(x)$ is known, though
the form given above, namely Eq.~(\ref{result}), is not particularly
transparent. 

For this mapping, though, based on the analysis for small $N$,  
we conjecture the following asymptotic expression for the 
probability density of the $N$--step exponent,
\beq
\label{cusp}
P( \lambda,N) = {N \over \pi} \frac{\exp (-N \vert \lambda 
- \L \vert)}{ [1- \exp (- 2N \vert \lambda - \L \vert)]^{1/2}} ~~~~ -\infty \leq \lambda \leq \ln 4.
\eqn
The main feature to be noted here is that there is a cusp at $\lambda = \L$,
which does not vanish even as $N \to \infty$ although
the range of $\lambda$ where it is significant decreases sharply 
with $N$. Outside this range, the function behaves essentially like
an exponential. Eq.~(\ref{cusp}) thus provides finite--size 
corrections to the
expressions derived earlier by Grassberger, Badii and Politi (see Eq.~4.9 in Ref.~ \cite{gbp}).

Both the results, namely Eq.~(\ref{result}) or Eq.~(\ref{cusp}) can 
be verified numerically and can be shown to hold to a high
level of accuracy. Shown in Fig.~2(a) are 
the (numerical) experimental distributions for the case of $N$ = 14,
which very closely matches both the exact (implicit) distribution,
Eq.~(\ref{result}), as well as the asymptotic result, 
Eq.~\ref{cusp} [Fig.~2(b)]. Evaluation of the former expression 
requires the determination of the roots of the corresponding
polynomial (we use the Newton--Raphson procedure).
Shown in Fig.~3 is the solution for the 
case of $N=3$, when the polynomial $G(x)$ is of order 7.
The divergences in the distribution $P(\lambda,3)$ occur as 
pairs of real roots merge.  For the specific case of the logistic
mapping, as $N$ increases the largest number of singularities 
accumulate near $\L = \ln 2$. The asymptotic form, Eq.~(\ref{cusp})
also gets progressively more accurate with increasing $N$ [see  Fig.~2(b)].

This form of the density is not restricted 
to the logistic mapping at $\alpha = 4$, but is also seen in
a number of mappings which have fully developed chaos.
Even in the logistic mapping, it occurs at {\it all}
parameter values corresponding to widening  crises \cite{ogy}, when the 
attractor is a rescaled image of the attractor at $\alpha = 4$.

We have also examined the dependence of the variance of the distribution 
on $N$. Typically, $ \sigma^2 \propto 1/N^{\gamma}$ for these
distributions since they narrow with increasing $N$, going, in the limit
of $N \to \infty$, to a $\delta$-function. For the case of a
Gaussian density, $\gamma $ = 1, while for the exponential
density the variance decreases more rapidly, and $\gamma $= 2. 
Our results for the variance are shown in Fig.~4.

\subsection{Intermittency} 
Intermittent dynamics is characterized by a long--range temporal 
persistence of correlations \cite{gh}, evidenced for example, by the 
existence of power--law dependence of a number of quantities 
on the parameters.  The question of the distribution of \lle 
for intermittent chaos has been explored previously by 
Benzi \etl \cite{benzi}, who showed that there are significant 
departures from the Gaussian distribution.

For the case of intermittency, the characteristic 
density of \lle appears to be a combination of a normal density  
and a stretched exponential tail, an example of which is shown in 
Fig.~1(c).
For $\lambda \leq \lambda_*$, the density is a Gaussian, 
while above $\lambda_*$, the dependence is \beq
P( \lambda, N) \approx \exp [-N^{\delta} (\lambda-\lambda_*)] ~~~~ \lambda
> \lambda_*.
\label{interb}
\eqn
Since the exponent $\delta < 1$, the exponential tail decays 
extremely slowly with $N$.  At $\lambda = \lambda_*$, there is 
a crossover between the Gaussian and the stretched exponential 
density, which results in a completely asymmetric density about the 
mean.  Similar distributions arise for all intermittent dynamical 
states, including the case of nonchaotic dynamics \cite{pmrl}.

One way of understanding the above (phenomenological) expression for 
the density is to note that in all intermittent dynamics, the motion 
switches between two types of states.  For each of these different
dynamical states, the local Lyapunov exponents have a Gaussian
distribution centered at different values of $\lambda$, and with
different amplitudes, and the stretched exponential behaviour 
interpolates between them. 

The example for which data is presented here in Fig.~1(c) and in 
Fig.~5 is the Type-I intermittency near the tangent bifurcation 
in the logistic mapping. This dynamics can be naturally separated
into laminar regions (that stay close to the incipient period--3 orbit)
and chaotic bursts. Finite--time Lyapunov exponents can be separately
computed for trajectories that stay entirely in the laminar phase and
entirely in the chaotic phase: these give the normal densities shown
by dotted lines in Fig.~5. Trajectory segments that visit both the
components of the intermittent motion contribute to the stretched
exponential tail; with increasing $N$, the purely chaotic component is 
more difficult to identify, since there are fewer segments that are
of duration longer than $N$. The stretched exponential tail thus decreases
with increasing $N$, and the distribution eventually collapses to 
a delta--function.

This behavior is generic at all intermittencies. For the mapping
\beq
x_{n+1} = x_n + c x_n^2,
\label{wang}
\eqn
which has been extensively studied within the thermodynamic formalism
in the context of Type-I intermittency \cite{wang}, we observe a similar
decomposition of the overall density of LLEs to a superposition of two
independent Gaussians with stretched exponential interpolation between
the two.  Furthermore, we have also examined a number of higher
dimensional maps and flows, and find that at all intermittent dynamics,
including the cases of forced systems or of nonchaotic dynamics
\cite{pmrl,rmp}, non--Gaussian stretched exponential tails are seen. An
example shown in Fig.~6 is the density $P(\lambda, 2048)$ for the
largest nonzero Lyapunov exponent in the Lorenz system (see the figure
caption for a definition of the dynamical system) with the parameters
chosen to correspond to intermittent dynamics.

The case of crisis--induced intermittency \cite{ott} (just beyond
widening crises, for instance) is similar, with the two independent
densities being, respectively, the exponential cusp, namely
Eq.~(\ref{cusp}) for the component for the pre--crisis chaotic
attractor, and a Gaussian density which corresponds to the widened
chaotic attractor. The density shown in Fig.~1(d) for the period--five
widening crisis in the logistic mapping can be analysed in a manner
very similar to the case illustrated in Fig.~5.

The variance for these distributions decrease somewhat faster that for
the Gaussian, namely $ \sigma^2 \propto 1/N^{\gamma}$ with $ \gamma
\equiv 2\delta > 1$.  With increasing $N$, the exponent however
changes, eventually reaching the Gaussian limit, $\gamma \approx 1$
(see Fig.~5).

\section{Summary and Discussion}

A major motivation for the present work has been the realization that
local Lyapunov exponent distributions have characteristic forms
depending on the nature of the attractors, and that these provide an
additional and important dynamical characterization of the chaotic
state of a system.
 
We have mainly focused on the cases of fully developed chaos, crises
and intermittencies---namely those attractors for which the density
shows a marked departure from the simple Gaussian form. This is 
indicative of significant finite--size corrections to the multifractal
formalism \cite{foot2}. All these cases 
show exponential tails albeit for different reasons \cite{time}. For
fully developed chaos in the logistic mapping we obtain the (in
principle) exact expression for the probability density as well as an
asymptotic approximate form. These densities are seen at all parameter
values corresponding to interior crises, and thus are quite common. 

The case of intermittency was analysed in detail, and the density seen
there was shown to arise from individual densities corresponding to
the different components of the motion (laminar and chaotic, say) with
the stretched exponential tail corresponding to interpolation between
these two.

We have verified the generality of our results for a number of other
dynamical systems, for example higher dimensional mappings such as the
H\'enon system or flows such as the Duffing oscillator, the forced
damped pendulum, the Lorenz equations \etc These densities are quickly
attained, and are maintained even as $N$ increases, and for fairly high
levels of additive noise \cite{noise}.  This is of particular relevance
when analyzing experimental data.

The logistic map was used here for illustration since exact results can
be obtained for at least one parameter value, but the probability
density for finite--time Lyapunov exponents in any system can be
obtained via Eq.~(\ref{result}) so long as the invariant measure is
known. In this regard, it is interesting to note that recently, Pingel
\etl \cite{ifp} have described a general inversion technique whereby a
class of 1--d maps having a prescribed invariant density can be
constructed. 

The characteristic forms for the density that we have described are
found in a variety of systems, including those where the dynamics is
not chaotic.  When a system is forced quasiperiodically, chaotic
attractors can be transformed to strange nonchaotic attractors
\cite{rmp}. These attractors are fractal, but the largest Lyapunov
exponent is zero or negative. Because of the spatial fractality,
though, for short times the local Lyapunov exponents can be positive,
and the distributions again fall into the classes that we have seen
here for chaotic systems \cite{trichy}. The phenomenon of high--stretch
tails also appears to be very general. In recent work Calvo and
Labastie \cite{calvo} have examined the distribution of \lles in a
conservative system, namely in a 19--particle cluster simulation study.
They observe that when the dynamics is intermittent \cite{cluster}, the
\le density has a stretched exponential tail.

{\it Local} analysis can be more revealing of the nature of the 
dynamics than global quantities, and this is an issue when predictability 
is of concern. For instance, in applications that aim to predict 
future behaviour based on time series data (for example in atmospheric 
sciences or in economics), atypical or extreme behaviour which 
contributes to the stretched tails is a serious bottleneck.
Since the largest Lyapunov exponent can be extracted from time--series 
data though standard techniques, examination of the finite time
distributions can give more insight into the dynamics than the
extraction of a single exponent.

We conclude with a few general comments. Except for fully developed
chaos when the distributions can be derived for all times, the present
study does not examine the case of very short times when the
distributions are atypical and it is not clear if they are stationary.
The characteristic behaviour becomes apparent for times that are not
too short, and persists thereafter.  Thus, local Lyapunov exponent
densities provide {\em quantitative} distinctions among different
chaotic attractors.  Exponential tails are characteristic of fully
developed chaos and intermittency: whether these are related to
analogous distributions that arise in turbulent flows \cite{tu2} or
stretched--exponential tails in relaxation phenomena is an interesting
open question.

\vskip1cm
\centerline{\bf ACKNOWLEDGMENT} 
We would
particularly like to thank Antonio Politi for numerous discussions 
and his generous advice.  We also thank Michael
Cross and Ed Ott for discussions, and the Max Planck Institute for
the Physics of Complex Systems, Dresden, where this work was completed, 
for hospitality.
This research was supported by a Grant
from the Department of Science and Technology, India. 


\centerline{Figure Captions}
\begin{itemize}
\item[Fig. 1] Characteristic probability densities that arise in
chaotic dynamical systems, corresponding to (a) ``typical'' chaos, (b)
fully--developed chaos, (c) intermittency, and (d) crisis-induced
intermittency. Numerical results are for $P(\lambda,N)$ for the
logistic map with (a) $\alpha$ = 3.7, (b) $\alpha$ = 4, and (c) $\alpha
= 1 + \sqrt 8 - 10^{-6}$, and (d) $\alpha = 3.7~447~104 $ . The
local Lyapunov exponents are calculated from $N$ step segments with
$10^{6}$ iterations. Different values of $N$ are chosen for the three
cases, in a) the Gaussian nature becomes apparent at $N$ around 10 or
so, the cusp in b) is already evident for $N$ around 8 and survives for
all $N$, while the asymmetry in the distribution c) and d) appears to
persist for all $N$. Note that for d), $N$ should be a multiple of the
periodicity of the window.

\item[Fig. 2] (a) Comparison, at $\alpha=4$, of the numerical results
(dots) for the density $P(\lambda,14)$, with the analytic expression,
Eq.~(\ref{result}), (solid line). (b) Numerical results (dots) compared
with the approximate density, Eq.~(\ref{cusp}) (solid line), for $N =
100$.

\item[Fig. 3] The real roots (solid line) and $P(\lambda,3)$ 
(dotted line) for the logistic map with $\alpha = 4$. The
singularities in the density  occur when  a pair of real roots
become complex. $P(\lambda,3)$ has been rescaled for clarity.

\item[Fig. 4] (a) The variance as a function of $N$ for ``typical'' 
chaos ($\Box$), fully--developed chaos ($\circ$), and crisis-induced 
intermittency ($\nabla$). The exponents characterizing
the decay are  $1.12, 1.95$, and $1.51$ respectively.
For intermittent chaos  (filled circles) there is a 
crossover: the exponents in different ranges of $N$ going from 
exponential limit, $1.85$ to the Gaussian limit $1.09$ at large $N$.

\item[Fig. 5] Near the intermittent transition, at [$\alpha =
1+ \sqrt(8)-10^{-6}$], where there are long crossovers, the two
components of the density $P( \lambda,300)$ (dotted lines)
 are compared with the total
density (solid line). For this state, $\lambda_p = -0.00015$ and 
$\lambda_* = 0.03$. Note that the amplitudes of the individual
densities have been appropriately scaled to depict clearly
the manner in which they contribute to the total density.

\item[Fig. 6] For the Lorenz equations : $ \dot{x} = a (y-x)$,
$\dot{y} = x (r - z)-y, \dot{z} = x y - b z $, where $a$, $b$ and $r$
are parameters, the density $P(\lambda, 2048)$ near the intermittency
transition at $a = 10, b = 8/3 \mbox{~and~} r = 166.8~801~548 $. Here
$\lambda$ is the largest nonzero Lyapunov exponent, and the integration
step size is 0.02 natural units.

\end{itemize}

\begin{thebibliography}{99}
\bibitem{foot1} The results of this section apply to continuous 
systems (flows) as well, and the formalism can be derived in an 
analogous manner.
\bibitem{corre}
The nature of these correlations can be seen in the plots of
$\lambda_N(i+1)$ versus $\lambda_N(i)$ where for the Gaussian case, points
will be uniformly distributed around the mean, while for the cases of
intermittency and fully developed chaos, points are heavily concentrated
in specific regions.
\bibitem{foot2} Such corrections can depend on the coordinates 
used (unlike the multifractal formalism itself).
\bibitem{time} It is important that the finite--time duration over which 
the local Lyapunov exponents are calculated not be too short. For $N$ below
10 in the discrete logistic mapping, the distributions are very atypical
and change very significantly with $N$. Furthermore, the larger $N$
distributions are more robust inasmuch as they are stable under smaller
sample sizes.
\bibitem{noise}
The effect of noise in the logistic map has been extensively studied
\cite{cfh}, and it is known that noise generally lowers 
the threshold for chaos: systems with additive noise have a larger
Lyapunov exponent for smaller nonlinearity.  As a consequence, the
finite--time Lyapunov exponents as well as their distributions can
change considerably under added noise.  We examined these effects by
adding noise in the dynamics, $x_{n+1} \to f(x_n) + \rho
\xi_n,$ where $\rho$ is the noise amplitude and the random variable
$\{\xi\}$ is $\delta$ correlated in time. The global structure of the
densities remains qualitatively unchanged for low noise amplitudes,
although with increased noise strengths, the distributions all tend to
the Gaussian.
\bibitem{cluster} Such distributions are seen in cluster dynamics
at energies corresponding to  solid--liquid coexistence;
this is a characteristic feature of phase change phenomena in finite systems.
\end{thebibliography}
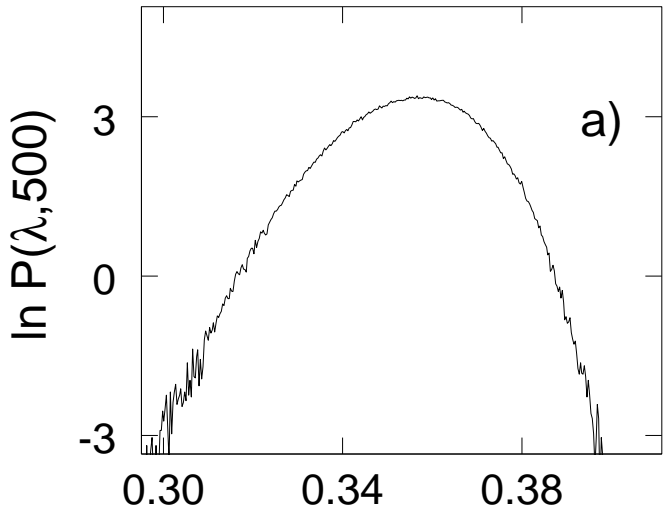
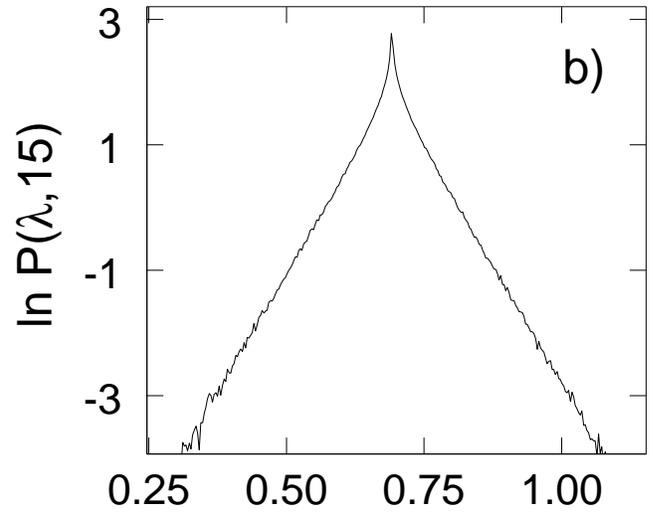
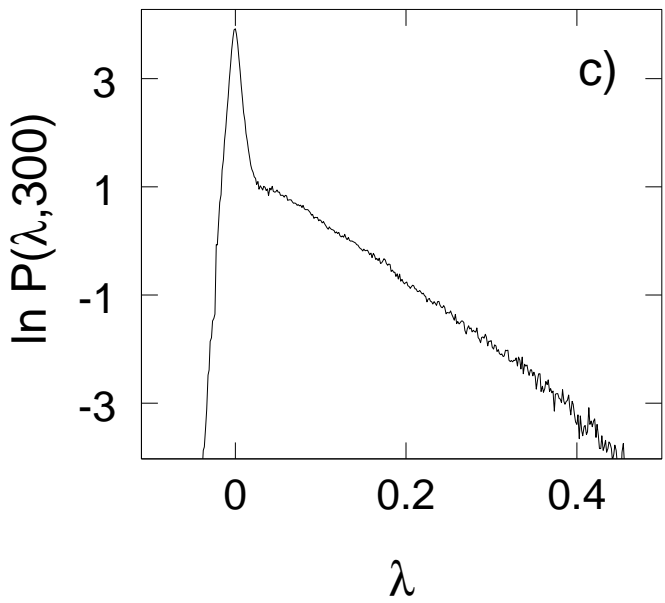
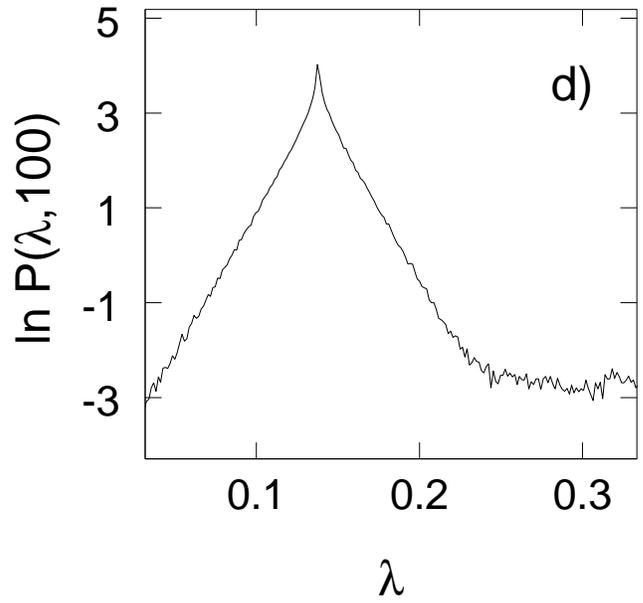
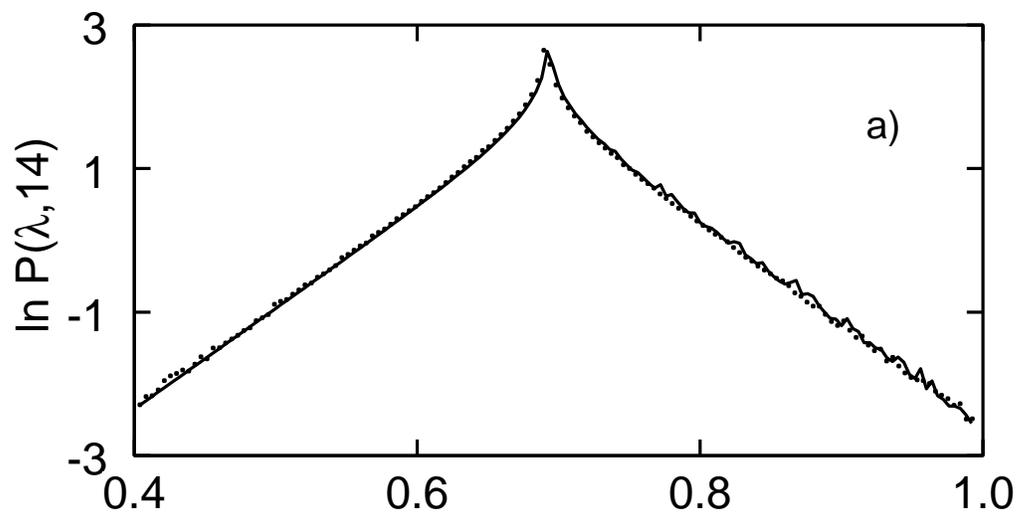
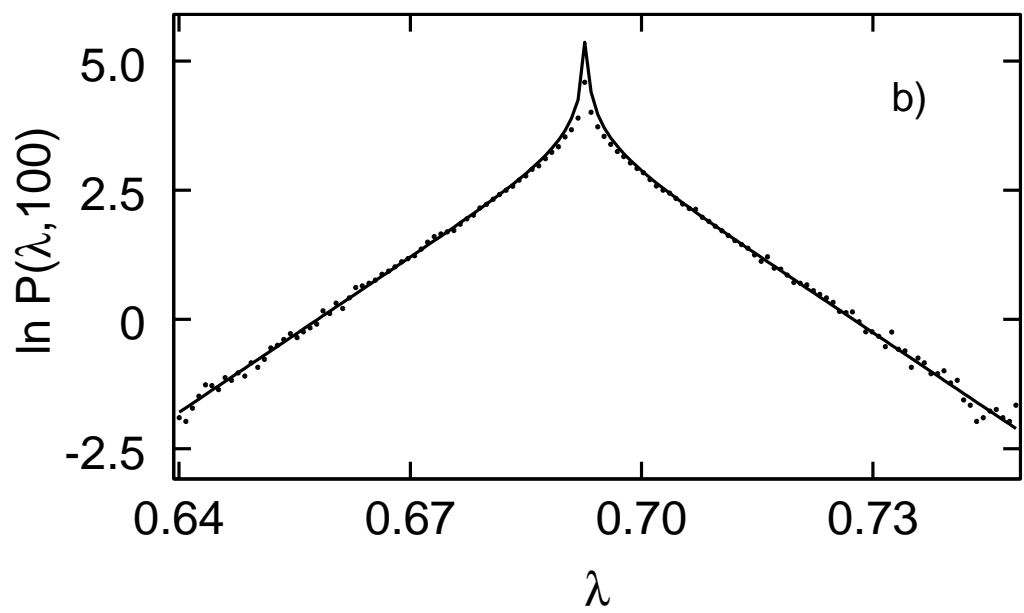
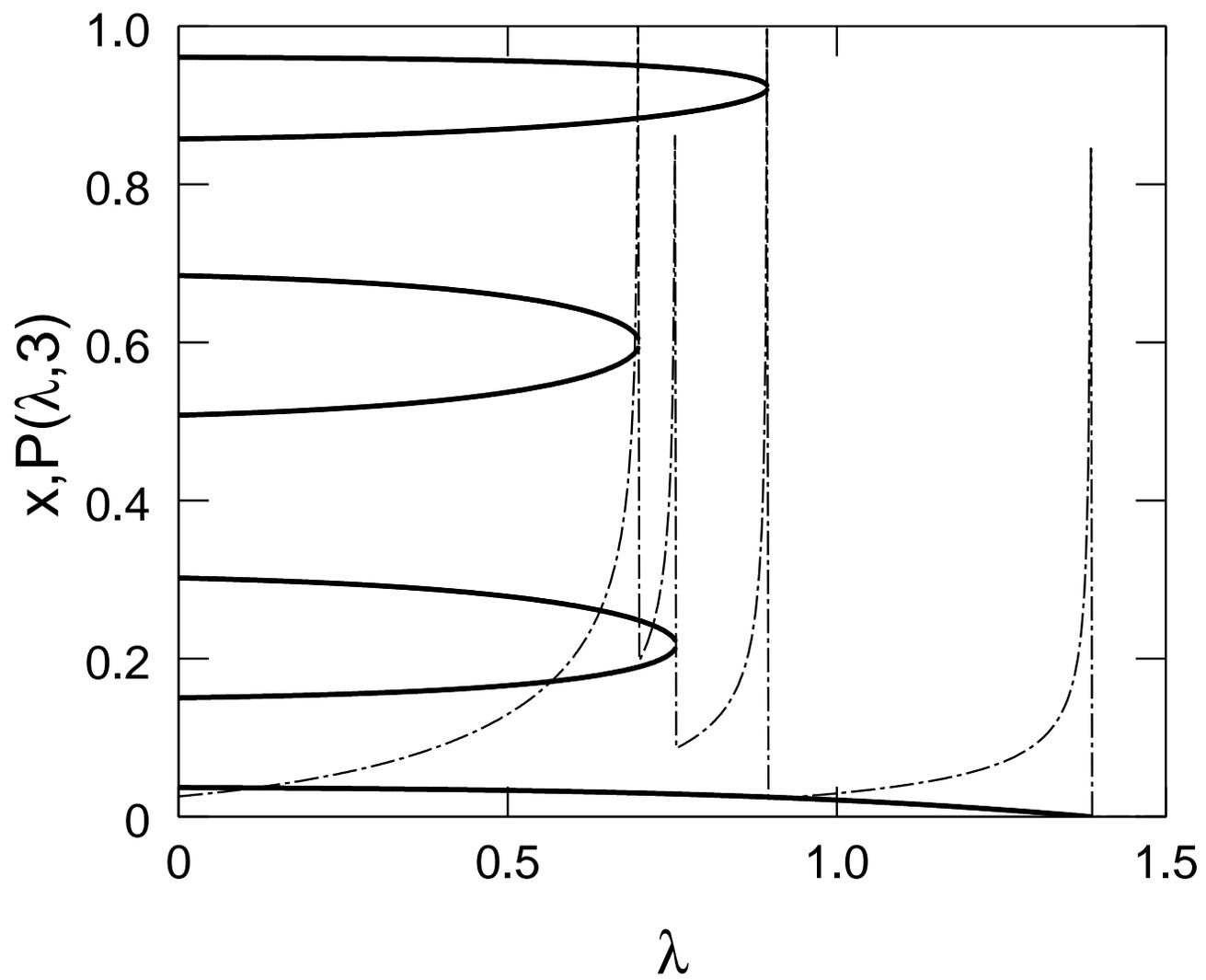
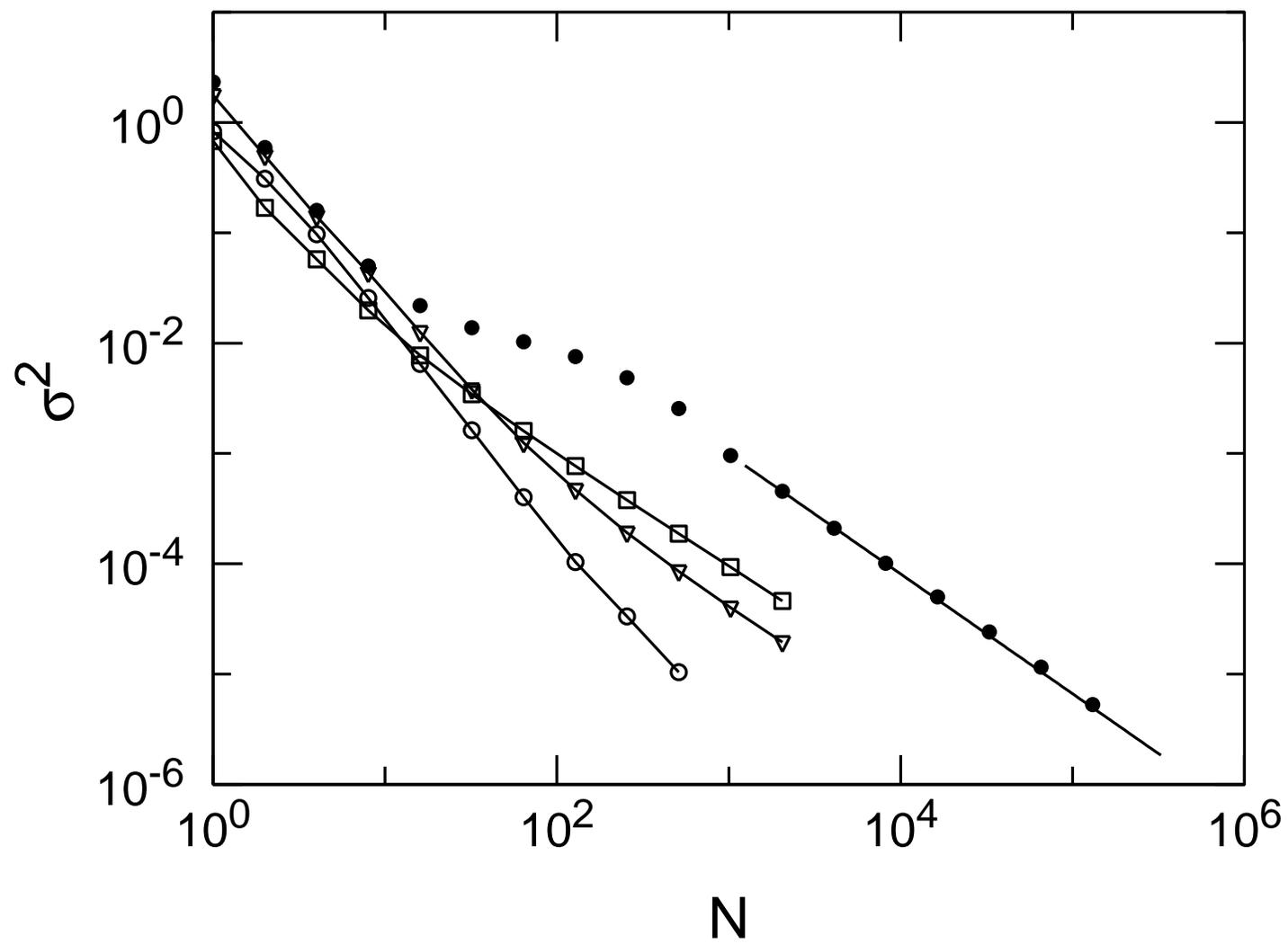
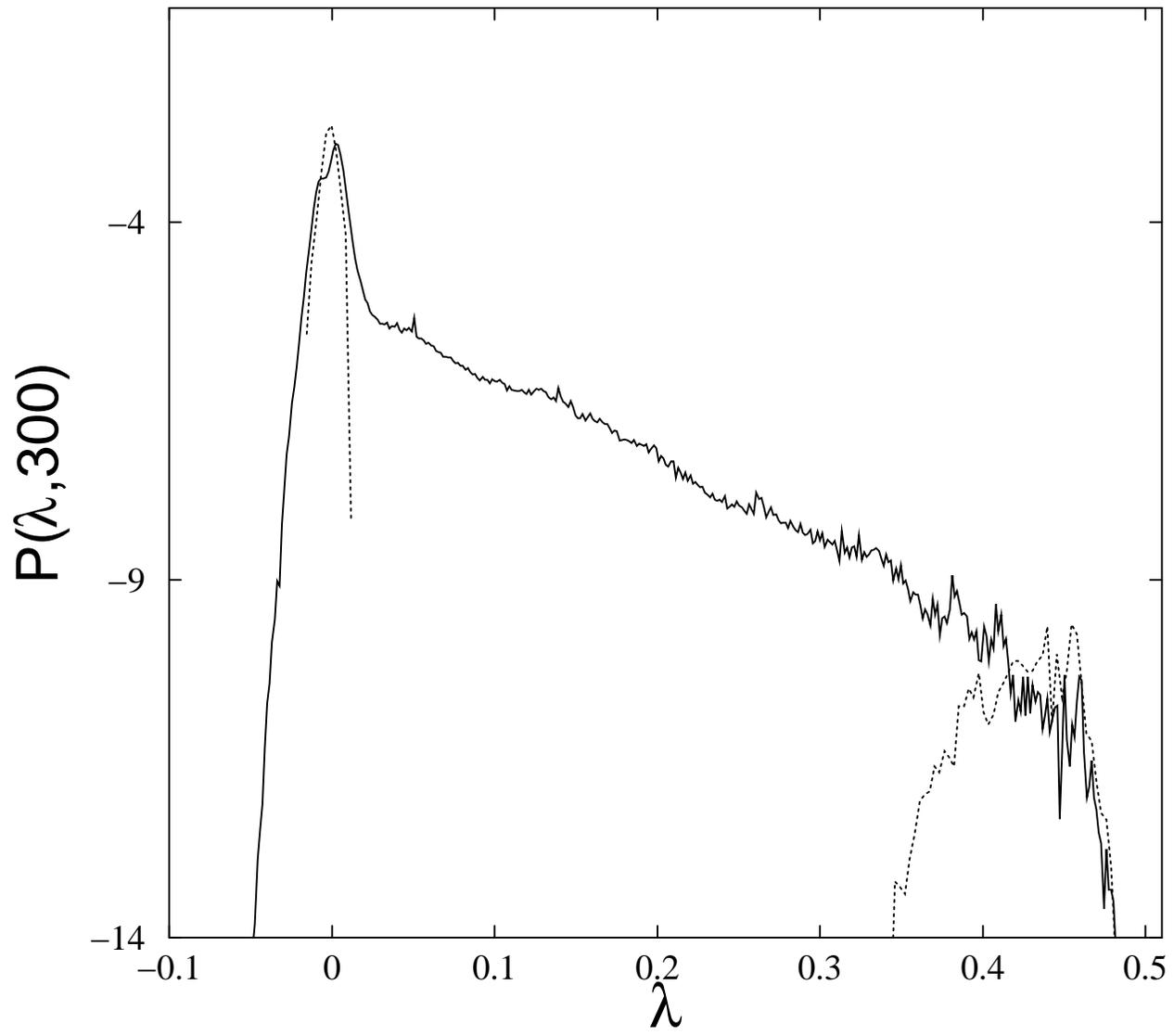
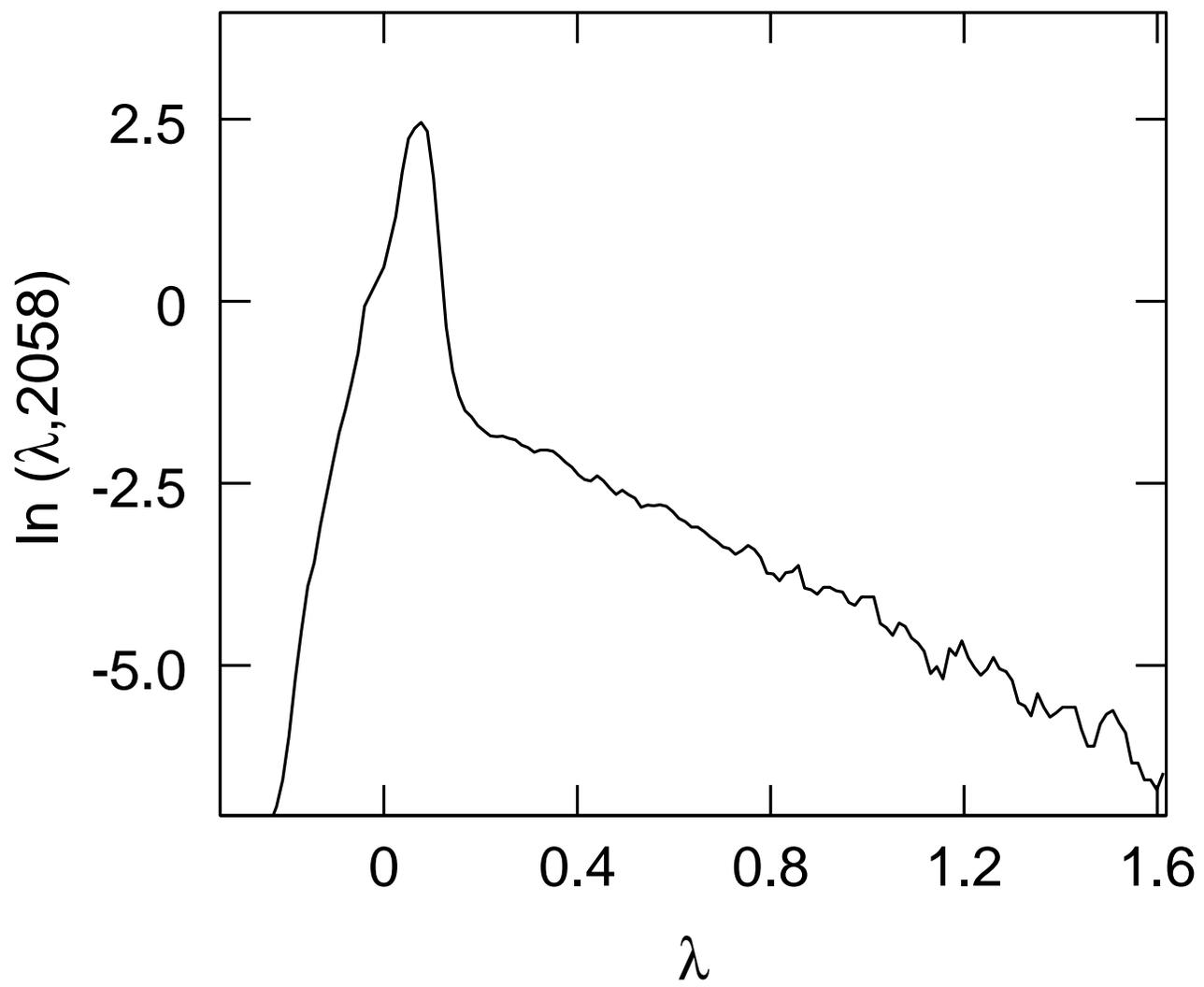
\end{document}